\documentclass[preprint,aps]{revtex4}

\usepackage{times}
\usepackage{graphicx}
\usepackage{dcolumn}
\usepackage{bm}
\usepackage{amsmath}

\begin{document}
\title{Integral equation for electrostatic waves generated by a point source
in a spatially homogeneous magnetized plasma}
\author{John J.~Podesta}
\affiliation{%
Space Science Institute, Boulder CO 80301}%
\email{jpodesta@solar.stanford.edu}




\date{\today}
\begin{abstract}
The electric field 
generated by a time varying point charge in a three-dimensional, unbounded, 
spatially homogeneous plasma with a uniform background magnetic field and a
uniform (static) flow velocity is studied in the electrostatic approximation
which is often valid in the near field.
For plasmas characterized by Maxwell distribution functions with isotropic 
temperatures, the linearized Vlasov-Poisson equations may be formulated in 
terms of an equivalent integral equation in the time domain.  The kernel
of the integral equation 
has a relatively simple mathematical form consisting of elementary 
functions such as exponential and trigonometric functions (sines and cosines), 
and contains no infinite sums of Bessel functions.  Consequently, the integral 
equation is amenable to numerical solutions and may be useful 
for the study of the impulse response of magnetized plasmas and, 
more generally, the response to arbitrary waveforms.  
\end{abstract}

\maketitle

\section{Introduction}

The study of plasma waves generated 
by an oscillating point charge or point dipole---fixed in 
space but oscillating in time---in a spatially homogeneous hot magnetized 
plasma is an important model problem 
that has been investigated both theoretically and 
experimentally since the 1960's \cite{Weitzner:1962, Kuehl:1962, Kuehl:1963,
Weitzner:1964, Chen:1964, Deering:1965, Fejer:1969, Fisher:1969, 
Fejer:1970, Fisher:1970, 
Schiff:1970, Fisher:1971, Singh_Gould:1971, Chasseriaux:1972, 
Tarstrup:1972, Baenziger:1972, Chen_Yen:1973, Singh_Gould:1973,
Kuehl:1973, Kuehl:1974a, Kuehl:1974b, 
Burrell:1975a, Burrell:1975b, Chasseriaux:1975, 
Ohnuma:1975, Ohnuma:1976, Bellan:1976, Simonutti:1976, 
Ohmori:1976, Ohnuma:1977a, Ohnuma:1977b, Ohnuma:1978a, Ohnuma:1978b,
Ohmori:1978, Ohnuma:1979a, Ohnuma:1979b, Lucks:1979, Lucks:1980, Beghin:1995}.  
Following the pioneering work of Landau in 1946 \cite{Landau:1946}, the 
fully kinetic self-consistent field 
approach described by the coupled Vlasov and Maxwell equations has been 
used to solve and study such systems, mainly through Fourier and Laplace
transform analysis.  
Experience with these techniques shows that even in the simplest spatially 
homogeneous systems,
the presence of ambient magnetic fields usually complicates the
mathematical analysis significantly.

Here we show that in the electrostatic approximation and for particle 
distribution functions with isotropic temperatures, an alternative approach 
based on solving an integral equation in the 
time domain provides a relatively simple mathematical formulation that 
contains no Bessel function series.  
Similar formulations may be possible for gyrotropic distributions with
anisotropic temperatures, but these are not considered here.
To the author's knowledge,
the integral equation presented here for the field of a point
source in a magnetized plasma has not appeared before in the literature.

Solutions of the linearized Vlasov-Maxwell or Vlasov-Poisson 
equations for specified charge and current distributions are of interest 
in many applications where 
the steady state solution for time harmonic forcing and the impulse 
response corresponding to a delta function forcing (in time) are 
usually of special interest.  In the electrostatic approximation 
and in the presence of a constant background magnetic field, the steady 
state solution for a point source can usually be expressed in terms of an 
infinite series of modified Bessel functions 
\cite{Kuehl:1973, Nakatani:1976, Ohnuma:1977a, Ohmori:1978, Piel:1984, Bonhomme:1994}.  
Here, it is shown that the
solutions for a point source with any prescribed time dependence
can be derived from an integral equation in which the kernel is free
of such Bessel function series.  Since the integral
equation formulation is amenable to numerical calculations, it can
be used to study the impulse response of magnetized plasmas as well
as the response to other arbitrary waveforms through direct calculation.  
In that respect, it has more flexibility and some other advantages compared to 
the Laplace transform approach.

\section{Statement of the problem}

Consider a spatially unbounded and homogeneous plasma with a background 
magnetic field $\bm B_0=B_0 \hat{\bm e}_z$, $B_0>0$, and an externally 
imposed point charge with charge density $q(t) \delta(\bm x)$, where 
$q(t)$ is a given function of time, $\delta(\bm x)=\delta(x)\delta(y)\delta(z)$ 
is the three dimensional delta function, $\delta(x)$  is the (one dimensional) 
Dirac delta function, and $(x,y,z)$ are the usual
orthogonal cartesian coordinates in three dimensional space with corresponding
unit vectors $\hat{\bm e}_x, \hat{\bm e}_y, \hat{\bm e}_z$.  
The function
$q(t)$ is assumed to vanish for $t\le 0$ and is continuous for $t\ge 0$.
Hence, by causality, the plasma response will vanish at $t=0$ or, in other words,
the initial conditions on the perturbed fields and the perturbed distribution
functions all vanish at $t=0$.  For $t>0$, the point charge excites 
plasma waves which propagate away from the origin and the 
problem is to compute the fields produced by this point source.

At equilibrium, the plasma is charge neutral and current free 
which is expressed by the relations
\begin{equation}
\sum_s n_s q_s = 0 \qquad \mbox{and}\qquad \sum_s n_s q_s\bm V_s = 0 
\end{equation}
where $n_s$, $q_s$ and $\bm V_s$ are the equilibrium number density, charge, 
and bulk flow velocity of particle species $s$, respectively.  
The equilibrium distribution functions are all assumed to be non-relativistic
convected Maxwell distributions with isotropic temperatures of the form
\begin{equation}
f_{0s}(v) = \frac{1}{(\pi v_s^2)^{3/2}} \exp\bigg[- \frac{|\bm v-\bm V_s|^2}
{v_s^2}\bigg],
\label{df}
\end{equation}
where $v_s=(2k_BT_s/m_s)^{1/2}$ is the thermal speed,
$k_B$ is Boltzmann's constant, $m_s$ is the particle mass, and $T_s$
is the kinetic temperature.  Note, however, that
the analysis below can readily be generalized to other distribution
functions with isotropic temperatures.  The Maxwell distribution (\ref{df}) 
has the convenient property
\begin{equation}
\nabla_{\bm v}f_{0s} = -\frac{2}{v_s^2} (\bm v-\bm V_s) f_{0s}(v)
\end{equation}
that shall be used below.

To linearize the Vlasov-Poisson equations, the first step is
to specify the equilibrium state.  In the presence of plasma flow in an 
arbitrary direction, the equilibrium solution of the collisionless Vlasov 
equation must satisfy $(\bm E_0 + \bm v \times \bm B_0) \cdot (\bm v-\bm V_s)=0$ 
which implies the existence of a static electric field 
$\bm E_0=-\bm V_s\times \bm B_0$.  For the solution to be self consistent,
the component of the velocity transverse to $\bm B_0$ must be the
same for each species, that is, $\bm V_s\times \bm B_0$ must 
be the same for each $s$.  The equilibrium state consisting of $\bm E_0$,
$\bm B_0$, and $f_0(\bm v)$ can now be used to linearize the Vlasov 
equation.  To simplify the presentation, the ion response 
shall be neglected from now on and only the electron contribution
shall be considered; because the susceptibilities of the different
particle species are additive, the ion response can easily be taken into 
account as shown at the end of the derivation.

The perturbed electric field $\bm E_1$ is assumed to be electrostatic
meaning that $\bm E_1 = -\nabla \phi$, where $\phi(\bm x, t)$ is the 
electrostatic potential, an approximation that is valid in the near field
\cite{Kuehl:1973, Fisher:1971},
at least over some range of physical parameters.  The precise range of 
validity of the electrostatic approximation for the problem under 
consideration is unknown and shall not be studied here. However,
for time harmonic forcing, the steady state solution can be computed 
numerically both with and without the electrostatic approximation, thus
providing a means of comparison.

In terms of the velocity variable $\bm u=\bm v-\bm V$, where the subscript 
on $\bm V=\bm V_e$ is omitted for convenience,  
the linearized Vlasov equation takes the form
\begin{equation}
\frac{\partial f_1}{\partial t} + (\bm u + \bm V)\cdot \nabla f_1 
-\frac{e}{m_e} (\bm u\times \bm B_0) \cdot \nabla_{\bm u} f_1 
=-\frac{e}{m_e}\nabla \phi \cdot \nabla_{\bm u} f_0,
\end{equation}
where $f_1(\bm x, \bm u, t)$ is the perturbed distribution function
defined such that the complete distribution function is $n_e(f_0 + f_1)$ 
with $f_0(\bm u)$ normalized to unity and $e>0$ the electronic charge.
Express the velocity vector in cyclindrical coordinates  as
\begin{equation}
\bm u = u_\perp \cos(\varphi) \hat{\bm e}_x + u_\perp \sin(\varphi) 
\hat{\bm e}_y + u_\parallel \hat{\bm e}_z.  
\label{v_cyl}
\end{equation}
After Fourier transformation with respect to the spatial variables, the
Vlasov equation becomes
\begin{equation}
\frac{\partial \widetilde f_1}{\partial t} + i\bm k\cdot(\bm u + \bm V) 
\widetilde f_1 
-\Omega_e  \frac{\partial \widetilde f_1}{\partial \varphi}
=\frac{2e\widetilde \phi}{v_e^2 m_e}i\bm k \cdot \bm u  f_0(u),
\label{V1}
\end{equation}
where $\widetilde f_1(\bm k, \bm u, t)$ and $\widetilde \phi(\bm k, t)$
are the respective Fourier transforms of $f_1(\bm x, \bm u, t)$ and
$\phi(\bm x, t)$, and $\Omega_e = -eB_0/m_e$ is the signed electron
cyclotron frequency $(\Omega_e<0)$.  The problem is to solve 
the Vlasov equation (\ref{V1}) for $t\ge 0$ together with Poisson's equation 
\begin{equation}
k^2\widetilde \phi(\bm k, t) = -\frac{n_e e}{\epsilon_0} \int \widetilde 
f_1(\bm k, \bm u, t) \, d\bm u + \frac{q(t)}{\epsilon_0}
\label{P1}
\end{equation}
subject to the initial conditions $\widetilde f_1(\bm k, \bm u, 0)=0$ and   
$\widetilde \phi(\bm k, 0)=0$, where $q(t)$ is a continuous function
such that $q(0)=0$.  Otherwise, the forcing function $q(t)$ is arbitrary.

\section{Formulation as an integral equation}

The goal is to reduce the initial value problem consisting of the Vlasov-Poisson 
equations (\ref{V1}) and (\ref{P1}) to an equivalent integral equation.
The result is as follows.  The potential $\widetilde \phi(\bm k, t)$ which solves the
Vlasov-Poisson system is also the solution of the Volterra integral 
equation of the second kind
\begin{equation}
\widetilde \phi(\bm k, t) + \omega_{pe}^2 \int_0^t \Gamma(\bm k, t- \tau) 
\widetilde \phi(\bm k, \tau) \, d\tau = \frac{q(t)}{\epsilon_0 k^2}, \qquad k\ne 0,
\label{IE}
\end{equation}
with kernel 
\begin{multline}
\Gamma(\bm k,\tau)=\tau \left[\frac{k_\perp^2}{k^2}\cdot
\frac{\sin(\Omega_e \tau)}{\Omega_e \tau}+\frac{k_\parallel^2}{k^2}\right] \\
\times   
\exp\bigg\{\! - \frac{k_\perp^2 v_e^2}{\Omega_e^2}\sin^2 \left(\frac{\Omega_e \tau}{2}\right)
-\left(\frac{k_\parallel v_e \tau}{2}\right)^{\! 2}-i\bm k\cdot\bm V\tau\bigg\}.
\label{ker}
\end{multline}
Note that the kernel is a convolution kernel as is required for a linear 
time invariant system.  Remarkably,  the kernel (\ref{ker}) does not 
contain the infinite sums of Bessel functions that 
occur in steady state solutions with time harmonic forcing.
The terms containing $\Omega_e$ represent the
magnetic field effects.  In the limit as $\Omega_e \rightarrow 0$,
the function (\ref{ker}) reduces to the correct kernel for a point charge
in an unmagnetized plasma    
\begin{equation}
\Gamma(\bm k,\tau)=\tau \exp\bigg\{\! - 
\left(\frac{k v_e \tau}{2}\right)^{\! 2}-i\bm k\cdot\bm V\tau\bigg\},
\label{ker0}
\end{equation}
where $k = (k_\perp^2 +k_\parallel^2)^{1/2}$.  Because the kernel (\ref{ker}) 
is smooth (continuously differentiable) and consists of compositions of
algebraic, exponential and trigonometric functions (sines and cosines),
it is convenient for purposes of numerical calculation.  When only the 
electron response is considered there are two inherent timescales which 
must be resolved, the electron cyclotron period $2\pi/\omega_{ce}$ and 
the inverse plasma frequency $2\pi/\omega_{pe}$.

Integral equation formulations of the
Vlasov-Poisson system in unmagnetized plasmas are well known 
\cite{Backus:1960a, Backus:1960b, Hayes:1961, Saenz:1965, Turski:1965, 
Lee_Soper:1968, Montgomery:1971, Colombo:1992, Podesta_POP:2010}.  
However, to the authors' knowledge, the integral equation
formulation for a magnetized plasma given by 
(\ref{IE}) and (\ref{ker}) is new.  Of course, the literature 
is vast, spanning more than 50 years, and it is possible
that this result may have been derived previously in work 
unknown to the present author.  
The remainder of this paper is devoted to a derivation of the integral 
equation (\ref{IE}) and the kernel (\ref{ker}).

\section{Solution of the Vlasov equation}\label{V}

Similar to (\ref{v_cyl}), let 
\begin{equation}
\bm k = k_\perp \cos(\theta) \hat{\bm e}_x + k_\perp \sin(\theta) 
\hat{\bm e}_y + k_\parallel \hat{\bm e}_z.
\end{equation}
Then 
\begin{equation}
\bm k \cdot \bm u= k_\perp u_\perp\cos(\varphi-\theta)+ k_\parallel u_\parallel. 
\end{equation}
The substitution 
\begin{equation}
\widetilde f_1=\psi \exp\bigg\{+i\frac{k_\perp u_\perp}{\Omega_e}
\sin(\varphi -\theta) -i(k_\parallel u_\parallel +\bm k\cdot\bm V)t\bigg\}
\label{sub}
\end{equation}
brings the Vlasov equation (\ref{V1}) into the form
\begin{equation}
\frac{\partial \psi}{\partial t}  -\Omega_e  \frac{\partial \psi}{\partial \varphi}
=\frac{2e}{v_e^2 m_e} \widetilde \phi(\bm k, t)f_0(u) i\bm k \cdot \bm u  
\exp\bigg\{-i\frac{k_\perp u_\perp}{\Omega_e}
\sin(\varphi -\theta) +i(k_\parallel u_\parallel +\bm k\cdot\bm V)t\bigg\}.
\label{V2}
\end{equation}
Consider the transformation from independent variables $t$ and $\varphi$
to a new set of independent variables $\eta$ and $\xi$ defined by
\begin{equation}
\left\{ \begin{array}{l}
\eta = {\textstyle \frac{1}{2}} (\varphi + \Omega_e t) \\
\xi = {\textstyle \frac{1}{2}} (\varphi - \Omega_e t). 
\end{array} \right. 
\end{equation}
It follows that 
\begin{equation}
\frac{\partial \psi}{\partial t}  -\Omega_e  \frac{\partial \psi}{\partial \varphi}
=-\Omega_e  \frac{\partial \psi}{\partial \xi}
\end{equation}
and, therefore, the solution of (\ref{V2}) is obtained by integrating both 
sides of that equation with respect to $\xi$ while holding $\eta$ fixed.
Because $t=0$ is equivalent to $\xi=\eta$, the integration is performed 
from the lower limit $\xi'=\eta$ to $\xi'=\xi$, where $\xi'$ is the 
variable of integration.  
To express the right-hand side of (\ref{V2}) in terms of $\eta$ and $\xi$,
substitute $\varphi = \eta+\xi$ and $t = (\eta-\xi)/\Omega_e$ so, for example,
\begin{equation}
\bm k \cdot \bm u= k_\perp u_\perp\cos(\eta +\xi-\theta)+ k_\parallel u_\parallel. 
\end{equation}
The integration of (\ref{V2}) implies
\begin{multline}
\psi(\eta, \xi) - \psi(\eta, \eta) = -\frac{2e}{v_e^2 m_e}
\int_\eta^\xi \frac{d\xi'}{\Omega_e} \, \widetilde \phi\Big(\bm k,
\frac{\eta-\xi'}{\Omega_e}\Big)f_0(u) e^{-i\bm k\cdot \bm V(\eta-\xi')/\Omega_e}  \\
\times i\big[k_\perp u_\perp\cos(\eta +\xi'-\theta)+ k_\parallel u_\parallel\big] 
\exp\bigg\{\! -i\, \frac{k_\perp u_\perp\sin(\eta +\xi'-\theta)+ 
k_\parallel u_\parallel (\xi'-\eta)}{\Omega_e}\bigg\}.
\end{multline}
By definition, $\psi(\eta, \eta)$ is equal to $\psi(t=0)$ which is equal to 
zero by virtue of the vanishing initial condition on $f_1$.  The change of 
variable $\tau=(\eta-\xi')/\Omega_e$, where $\eta$ is a constant, yields
\begin{multline}
\psi(\eta, \xi)  =\frac{2e}{v_e^2 m_e}
\int_0^t d\tau \, \widetilde \phi(\bm k,\tau)f_0(u)i\big[k_\perp u_\perp\cos(2\eta -\theta - \Omega_e\tau)
+ k_\parallel u_\parallel\big] \\
\exp\bigg\{ -i\, \frac{k_\perp u_\perp}{\Omega_e}\sin(2\eta -\theta - \Omega_e\tau)+ 
i(k_\parallel u_\parallel +\bm k\cdot \bm V)\tau \bigg\}.
\end{multline}
Because $\eta$ is held constant inside the integrand, if one makes the 
substitution $\eta = (\varphi +\Omega_e t)/2$  inside the integrand with
both $\varphi$ and $t$ held constant during the integration,
then the result of the integration is the same as if the substitution 
$\eta = (\varphi +\Omega_e t)/2$ were made after the integration.
Therefore, the preceding equation is equivalent to
\begin{multline}
\psi(\bm k, \bm u, t)  =\frac{2e}{v_e^2 m_e}
\int_0^t d\tau \, \widetilde \phi(\bm k,\tau)f_0(u)i\big\{k_\perp u_\perp
\cos[\varphi -\theta + \Omega_e(t-\tau)]
+ k_\parallel u_\parallel\big\} \\
\exp\bigg\{ -i\, \frac{k_\perp u_\perp}{\Omega_e}\sin[\varphi -\theta + \Omega_e(t-\tau)]+ 
i(k_\parallel u_\parallel +\bm k\cdot \bm V)\tau \bigg\}.
\end{multline}
It can be verified by direct calculation that this is the solution 
of the partial differential equation (\ref{V2}). Thus, from (\ref{sub}), 
the solution of the Vlasov equation (\ref{V1}) is
\begin{multline}
\widetilde f_1(\bm k, \bm u, t)  =\frac{2e}{v_e^2 m_e}
\int_0^t d\tau \, \widetilde \phi(\bm k,\tau)f_0(u)i\big\{k_\perp u_\perp
\cos[\varphi -\theta + \Omega_e(t-\tau)]
+ k_\parallel u_\parallel\big\} \\
\exp\bigg[\! -i\, \frac{k_\perp u_\perp}{\Omega_e}\Big\{\sin[\varphi -\theta 
+ \Omega_e(t-\tau)]-\sin(\varphi -\theta)\Big\} -
i(k_\parallel u_\parallel +\bm k\cdot \bm V)(t-\tau) \bigg]
\end{multline}
or, equivalently, 
\begin{multline}
\widetilde f_1(\bm k, \bm u, t)  =\frac{2e}{v_e^2 m_e}
\int_0^t d\tau \, \widetilde \phi(\bm k,t-\tau)f_0(u)i\big[k_\perp u_\perp
\cos(\varphi -\theta + \Omega_e \tau)
+ k_\parallel u_\parallel\big] \\
\exp\bigg\{\! -i\, \frac{k_\perp u_\perp}{\Omega_e}\big[\sin(\varphi -\theta 
+ \Omega_e \tau)-\sin(\varphi -\theta)\big] -
i(k_\parallel u_\parallel +\bm k\cdot \bm V)\tau \bigg\}
\label{f0}
\end{multline}
This result is now inserted into Poisson's equation 
(\ref{P1}) and the integration over velocity space is performed.

\section{Velocity space integrals}\label{I}

The integration of $\widetilde f_1$ with respect to $\bm u$ is performed 
in cylindrical coordinates. To facilitate the calculation, it is 
expedient to write (\ref{f0}) in the form
\begin{multline}
\widetilde f_1(\bm k, \bm u, t)  =-\frac{2e}{v_e^2 m_e}
\int_0^t d\tau \, \widetilde \phi(\bm k,t-\tau)f_0(u)e^{-i\bm k\cdot\bm V\tau}  \\
\times \frac{\partial \,}{\partial \tau} \exp\bigg\{\! -i\, \frac{k_\perp u_\perp}
{\Omega_e}\big[\sin(\varphi -\theta + \Omega_e \tau)-\sin(\varphi -\theta)\big] -
ik_\parallel u_\parallel \tau \bigg\}
\label{f1}
\end{multline}
The integal with respect to $\varphi$ is readily computed using the 
Bessel function series
\begin{equation}
e^{iz\sin\phi} = \sum_{n=-\infty}^\infty e^{in\phi} J_n(z),
\end{equation}
where $J_n(z)$ is the Bessel function of the first kind of order $n$ and 
argument $z$, and $z$ is an arbitrary complex number \cite{Abramowitz_Stegun}.  
It follows from this series expansion that 
\begin{equation}
\exp\bigg[ i\,\frac{k_\perp u_\perp}{\Omega_e}\sin(\varphi -\theta)\bigg]
 = \sum_{m=-\infty}^\infty e^{im(\varphi -\theta)} J_m\Big(
\frac{k_\perp u_\perp}{\Omega_e}\Big)
\label{s1}
\end{equation}
and 
\begin{equation}
\exp\bigg[ i\,\frac{k_\perp u_\perp}{\Omega_e}\sin(\varphi -\theta+ 
\Omega_e \tau)\bigg] = \sum_{n=-\infty}^\infty e^{in(\varphi -\theta+ 
\Omega_e \tau)} J_n\Big(\frac{k_\perp u_\perp}{\Omega_e}\Big).
\label{s2}
\end{equation}
The orthogonality relations for the complex exponentials then yield
\begin{multline}
\int_0^{2\pi} d\varphi \, \exp\bigg\{\! -i\, \frac{k_\perp u_\perp}
{\Omega_e}\big[\sin(\varphi -\theta + \Omega_e \tau)-\sin(\varphi -\theta)\big] 
\bigg\} \\
= 2\pi \sum_{n=-\infty}^\infty \Big[J_{n}\Big(\frac{k_\perp u_\perp}{\Omega_e}
\Big)\Big]^{\! 2}e^{in \Omega_e \tau}.
\end{multline}
The series on the right-hand side is a special case of the Fourier series
\begin{equation}
\sum_{n=-\infty}^\infty \big[J_n(z)\big]^2 e^{in\phi} = J_0
\Big(2z\sin\frac{\phi}{2}\Big)
\end{equation}
with Fourier coefficients given by \cite[][page 48, equation 13]{Erdelyi:1953}
\begin{equation}
\frac{1}{2\pi}\int_{-\pi}^{\pi} J_0\Big(2z\sin\frac{\phi}{2}\Big)e^{-in\phi}
\, d\phi = \big[J_n(z)\big]^2.
\end{equation}
Hence, 
\begin{multline}
\int_0^{2\pi} d\varphi \, \exp\bigg\{\! -i\, \frac{k_\perp u_\perp}
{\Omega_e}\big[\sin(\varphi -\theta + \Omega_e \tau)-\sin(\varphi -\theta)\big]
\bigg\} \\
= 2\pi J_0\left(\frac{2k_\perp u_\perp}{\Omega_e}\sin \frac{\Omega_e \tau}{2}\right)
\end{multline}
and, from (\ref{f1}), 
\begin{multline}
\int_0^{2\pi} d\varphi \, \widetilde f_1(\bm k, \bm u, t)=-\frac{4\pi e}{v_e^2 m_e}
\int_0^t d\tau \, \widetilde \phi(\bm k,t-\tau)f_0(u)e^{-i\bm k\cdot\bm V\tau}  \\
\times \frac{\partial \,}{\partial \tau} \bigg\{
J_0\left(\frac{2k_\perp u_\perp}{\Omega_e}\sin \frac{\Omega_e \tau}{2}\right)
e^{-ik_\parallel u_\parallel \tau} \bigg\}
\end{multline}

It remains to perform the integration  with respect to $u_\perp$ and
$u_\parallel$.  The integration with respect to $u_\perp$ is accomplished
by means of Weber's first exponential integral \cite[][page 393]{Watson:1944} 
\begin{equation} 
\frac{1}{\pi v_e^2}
\int_0^\infty u_\perp e^{u_\perp^2/v_e^2} J_0(a u_\perp)
\, du_\perp  = \frac{1}{2\pi} \exp\bigg\{-\frac{v_e^2a^2}{4}\bigg\}.
\label{formula}
\end{equation}
The application of (\ref{formula}) yields
\begin{multline}
\int_0^\infty u_\perp du_\perp \int_0^{2\pi} d\varphi \, 
\widetilde f_1(\bm k, \bm u, t)=-\frac{2e}{v_e^2 m_e}
\int_0^t d\tau \, \widetilde \phi(\bm k,t-\tau)f_0(u_\parallel)
e^{-i\bm k\cdot\bm V\tau}  \\
\times \frac{\partial \,}{\partial \tau} \bigg\{
\exp\left(\! - \frac{k_\perp^2 v_e^2}{\Omega_e^2}\sin^2 \frac{\Omega_e \tau}{2}\right)
e^{-ik_\parallel u_\parallel \tau} \bigg\}.
\end{multline}
And the integration with respect to $u_\parallel$ now gives
\begin{multline}
\int_0^\infty u_\perp du_\perp \int_{-\infty}^\infty du_\parallel 
\int_0^{2\pi} d\varphi \, \widetilde f_1(\bm k, \bm u, t)=-\frac{2e}{v_e^2 m_e}
\int_0^t d\tau \, \widetilde \phi(\bm k,t-\tau)
e^{-i\bm k\cdot\bm V\tau}  \\
\times \frac{\partial \,}{\partial \tau} 
\exp\bigg\{\! - \frac{k_\perp^2 v_e^2}{\Omega_e^2}\sin^2 \left(\frac{\Omega_e \tau}{2}\right)
-\left(\frac{k_\parallel v_e \tau}{2}\right)^{\! 2}\bigg\}.
\end{multline}
Differentiation with respect to $\tau$ yields the final result 
\begin{equation}
\int \widetilde f_1(\bm k, \bm u,t) \, d\bm u  =\frac{k^2e}{m_e}
\int_0^t d\tau\, \Gamma(\bm k,\tau)\widetilde \phi(\bm k, t-\tau),
\end{equation}
where $\Gamma(\bm k, \tau)$ is the kernel (\ref{ker}).  Substitution 
of this result into Poisson's equation (\ref{P1}) yields the 
integral equation (\ref{IE}).  This completes the derivation of
(\ref{IE}) and (\ref{ker}).

\section{Integral equation for multi-species plasmas}

Different types of ions and other charged particle species can be 
taken into account in a straightforward manner.
The Vlasov equation for particle species $s$ takes the form
\begin{equation}
\frac{\partial \widetilde f_{1s}}{\partial t} + i\bm k\cdot(\bm u + \bm V_s) 
\widetilde f_{1s} 
-\Omega_s  \frac{\partial \widetilde f_{1s}}{\partial \varphi}
=-\frac{2q_s\widetilde \phi}{v_s^2 m_s}i\bm k \cdot \bm u  f_{0s}(u),
\label{Vx}
\end{equation}
where $\widetilde f_{1s}(\bm k, \bm u, t)$ is the Fourier transform of 
$f_{1s}(\bm x, \bm u, t)$ and $\Omega_s = q_sB_0/m_s$ is the signed 
cyclotron frequency of species $s$.  The effects of the different species are
additive in Poisson's equation which becomes
\begin{equation}
k^2\widetilde \phi(\bm k, t) = \sum_s \frac{n_s q_s}{\epsilon_0} \int \widetilde 
f_{1s}(\bm k, \bm u, t) \, d\bm u + \frac{q(t)}{\epsilon_0}.
\label{Px}
\end{equation}
Subject to the initial conditions $\widetilde f_{1s}(\bm k, \bm u, 0)=0$ and   
$\widetilde \phi(\bm k, 0)=0$, the Vlasov equation (\ref{Vx}) is solved 
as described in Section \ref{V} and the velocity space integrals are
computed as in Section \ref{I} with the result
\begin{equation}
\widetilde \phi(\bm k, t) + \int_0^t \Gamma(\bm k, t- \tau) 
\widetilde \phi(\bm k, \tau) \, d\tau = \frac{q(t)}{\epsilon_0 k^2}, \qquad k\ne 0,
\label{IEx}
\end{equation}
where
\begin{multline}
\Gamma(\bm k,\tau)= \sum_s \omega_{ps}^2 \tau \left[\frac{k_\perp^2}{k^2}\cdot
\frac{\sin(\Omega_s \tau)}{\Omega_s \tau}+\frac{k_\parallel^2}{k^2}\right] \\
\times   
\exp\bigg\{\! - \frac{k_\perp^2 v_s^2}{\Omega_s^2}\sin^2 \left(\frac{\Omega_s \tau}{2}\right)
-\left(\frac{k_\parallel v_s \tau}{2}\right)^{\! 2}-i\bm k\cdot\bm V_s\tau\bigg\}
\label{kerx}
\end{multline}
and $\omega_{ps}^2=n_s q_s^2/\epsilon_0 m_s$.  Even though the integral 
equation formulation of the Vlasov-Poisson system is exact, the calculation 
of numerical solutions in multi-species plasmas is difficult in practice 
because of the disparity between ion and electron timescales.

\section{Conclusions}  

It has been shown that
the electric field generated by a time varying point charge $q(t)\delta(\bm x)$ in a 
magnetized plasma can be obtained as the solution of an integral equation 
and, in the case of a convecting Maxwellian plasma with an isotropic pressure 
distribution, that the kernel may be expressed in terms of elementary 
functions.  Numerical solutions of this integral equation can be used to 
study the response of the plasma to various types of forcing $q(t)$ and 
thus it provides an
alternative approach to exact analytic theories and their approximations 
which are often cumbersome for the analysis of magnetized plasmas.  
For example, the impulse response of a
magnetized plasma can be studied by using pulsed sources such as 
$q(t)=At\exp(-t/t_0)$ or $q(t)=At\exp(-t^2/2t_0^2)$.
These functions both rise to a maximum at $t=t_0$ and then decline to 
zero; the risetime $t_0$ is adjustable, of course, for the application 
of interest.  The integral equation can also be used to study 
wave excitation and propagation for sources with
arbitrary waveforms such that $q(t)$ is continuous and satisfies $q(0)=0$.  
The integral equation, therefore, should be of practical value.

Concerning the avoidance of infinite series of Bessel functions
in studies of homogeneous magnetized plasmas, the recent work by Qin 
{\it et al.}  \cite{Qin:2007} should be mentioned.
Textbook derivations of the
electromagnetic susceptibility of homogeneous, magnetized, gyrotropic plasmas
usually employ expansions in infinite series of Bessel functions to perform
the integration over the particle gyro-phase \cite[see, for example,
chapter 10 in Ref.][]{Stix:1992}.  Qin {\it et al.}  \cite{Qin:2007} have
shown how this integration may be carried out without the use
of Bessel function expansions.  While this is an important mathematical
development, the resulting form of the susceptibility
tensor derived by Qin {\it et al.} still contains integrations over
the variables $p_\perp$ and $p_\parallel$ which have not been carried out.
In the case of Maxwell or bi-Maxwell velocity distribution functions,
the remaining integrations over
$p_\perp$ and $p_\parallel$ inevitably give rise
to infinite series of Bessel functions and lead to the well known
expressions given, for example, by Stix \cite{Stix:1992}.  Consequently,
for Maxwellian plasmas, the expressions for the susceptibility tensor 
derived by Qin {\it et al.} 
have postponed but not eliminated the practical need for such series 
expansions in numerical calculations of the hot plasma dielectric tensor.


\begin{acknowledgments}
This work was supported by
the NASA Solar and Heliospheric Physics Program and by the 
NSF Shine Program.
\end{acknowledgments}

\bigskip



\hyphenation{Post-Script Sprin-ger}

\end{document}